# Landau theory of compressible magnets near a quantum critical point

Gillian A Gehring[1] and Mahrous R. Ahmed[2]

[1] Department of Physics and Astronomy, University of Sheffield, Sheffield S3 7RH, UK

[2] Department of Physics, University of Sohag, Sohag 82534, Egypt.

**Abstract**

Landau theory is used to investigate the behaviour of a metallic magnet driven towards a quantum critical point by the application of pressure. The observed dependence of the transition temperature with pressure is used to show that the coupling of the magnetic order to the lattice diverges as the quantum critical point is approached. This means that a first order transition will occur in magnets (both ferromagnets and antiferromagnets) because of the coupling to the lattice. The Landau equations are solved numerically without further approximations. There are other mechanisms that can cause a first order transition so the significance of this work is that it will enable us to determine the extent to which any particular first order transition is driven by coupling to the lattice or if other causes are responsible.

**I Introduction**

A quantum critical point (QCP) is approached by applying pressure in a number of conducting ferromagnetic and antiferromagnetic metals [1,2,3]. This involves a transition from a magnetic metal to a phase that is a strongly correlated Fermi Liquid. In some cases a superconducting phase appears as the QCP is approached, however in a number of materials a first order transition occurs just before the QCP is reached [4]. The observed dependence of $T_c$ on pressure necessarily means that the magnetic energy is coupled to the lattice. A first order phase transition occurs if this coupling exceeds a critical value: this is inevitable if the pressure derivative of $T_c$ diverges at the QCP as occurs for metallic materials [5] for which a power law dependence is found for the dependence of the transition temperature on pressure $T_c(p) = T_0 \left(1 - p/p_0\right)^\eta$ where $\eta < 1$. There are good theoretical arguments for expecting that $\eta = 3/4$ [6] so we shall use this value in what follows. The temperature and magnetic field behaviour of many of the weak metallic magnets e.g. $ZrZn_2$ are described very well by Landau theory over a wide parameter range [7] so it is useful to use this approach to get the whole phase diagram.

The physics of the problem is understood from the fact the magnetostriction will expand the lattice in the ordered phase because the transition temperature falls when the material is compressed. This leads naturally to a first order phase transition because we can reach a pressure where the transition temperature in the paramagnetic phase is above the ambient temperature but because of magnetostriction the strain in the ordered phase is reduced so the transition temperature drops sufficiently so that the ordered phase is stable. Thus there is a mixed phase region consisting of nonmagnetic and magnetic regions. The magnetisation is also enhanced in the second order regime because the magnetic energy is enhanced below the ordering temperature due to the magnetostriction.

Both the magnetisation and hence the magnetostriction vanish at the critical temperature of a second order transition and so the transition temperature may be defined in terms of the strain, $T_c(\varepsilon) = T_0\left(1 - \varepsilon/\varepsilon_0\right)^{3/4}$ where $\varepsilon_0 = p_0/K$ and $K$ is the elastic modulus. This is the expression used to write the strain in the material in terms of the strain due to the applied pressure which is present in the paramagnetic phase, $\varepsilon_s = p/K$ and the additional strain due to magnetostriction. This generates a coupling between the square of the order parameter and the magnetostrictive strain.

## II Linear theory

The Landau free energy of a magnetic material, including the hydrostatic pressure, for which the transition temperature, $T_c$ is a function of the strain $\varepsilon$, is written as follows:

$$F(M,\varepsilon; p,T) = \frac{A}{2}(T - T_c(\varepsilon))\mathbf{M}^2 + \frac{B}{4}(\mathbf{M}^2)^2 + \frac{C}{6}(\mathbf{M}^2)^3 + O(M^8) + \frac{K\varepsilon^2}{2} - p\varepsilon \qquad (1)$$

In the paramagnetic phase the equilibrium value of the strain is given by $\varepsilon_s = p/K$. In the linearised theory $T_c$ is expanded about $\varepsilon_s$ to first order, $T_c(\varepsilon) = T_c(\varepsilon_s) + \left.\frac{\partial T_c(\varepsilon)}{\partial \varepsilon}\right|_{\varepsilon=\varepsilon_s} (\varepsilon - \varepsilon_s)$.

The free energy is minimised with respect to $(\varepsilon - \varepsilon_s)$ and this produces a renormalisation of the fourth order coefficient,

$$B' = B\left(1 - \frac{2\zeta^2}{KB}\right) \quad \text{where} \quad \zeta = \frac{A}{2}\left.\frac{\partial T_c(\varepsilon)}{\partial \varepsilon}\right|_{\varepsilon=\varepsilon_s} \qquad (2)$$

The tricritical point occurs where $B'$ goes to zero. The enhancement of the magnetisation in the ordered phase, mentioned earlier occurs because of the reduced value of $B$. The magnetisation is evaluated as a function of temperature for the second order and tricritical regimes in this model in order to compare it with the self consistent theory described below. Although this theory does give a first order region the value of the magnetisation in the ordered phase is unreasonably high.

**III Self consistent theory**

In the case that the reduction in the transition temperature follows a power law as a function of strain or pressure as is seen in metallic materials a more complete theory can be used [5].

The minimum of the free energy (1) is given by the solution of the two equations,

$$\frac{\partial F(M,\varepsilon;p,T)}{\partial M} = 0 = M\left[A(T-T_c(\varepsilon)) + B\mathbf{M}^2 + O(M^4)\right] \quad (3)$$

$$\frac{\partial F(M,\varepsilon;p,T)}{\partial \varepsilon} = 0 = -\frac{A}{2}\frac{\partial T_c(\varepsilon)}{\partial \varepsilon}\mathbf{M}^2 + K\varepsilon - p \; ; \quad (4)$$

the solutions are: $M=0$ or $M^2 = \frac{A(T_c(\varepsilon)-T)}{B}$ \quad (5)

$$\varepsilon = \frac{p}{K} + \frac{A^2}{2BK}(T_c(\varepsilon)-T)\frac{\partial T_c}{\partial \varepsilon} \; . \quad (6)$$

We solve these equations numerically using $T_c(\varepsilon) = T_0\left(1 - \varepsilon/\varepsilon_0\right)^{3/4}$. In the nonmagnetic phase the strain takes the temperature independent value, $\varepsilon = \varepsilon_s = p/K$. The parameters were chosen such that the tricritical point occurs for $\varepsilon_s/\varepsilon_0 = 0.9$. In fig. 1 we show the dependence of the strain with the magnetisation as calculated from equan. 3. The magnetisation is divided by a normalisation factor, $M_0^2 = \dfrac{AT_0}{B}$ which is the value of the saturation magnetisation when $T_c$ is independent of strain. When $M=0$ the strain takes the value $\varepsilon = \varepsilon_s$ but for finite magnetisation the strain is always below this value in some cases by a very significant amount. The curves are only drawn for the range of magnetization that actually occurs for that value of the applied strain. Since the transition temperature depends on the strain we can combine the results shown in fig. 1 to calculate the dependence of the transition temperature on the magnetisation and this is plotted in fig. 2. This shows the expected effect that $T_c$ increases with the magnetisation; but we see from equation (3) that an increase in $T_c$ leads in turn to an increase in the magnetisation. It is this positive feedback that leads to a first order transition. Near to the QCP the transition temperature is a very sensitive function of the strain and so the linear theory is valid only when the magnetisation is very small and hence this theory is valid over a very narrow temperature range.

This is shown in fig. 3 which shows the temperature dependence of the strain obtained from the self consistent solution of equation (4). There are three different types of curves shown; in all cases $\varepsilon < \varepsilon_s$ because $\dfrac{\partial T_c(\varepsilon)}{\partial \varepsilon} < 0$ . For $\varepsilon_s/\varepsilon_0 <0.9$, in the second order region, the curve of $\varepsilon/\varepsilon_0$ approaches the value of $\varepsilon_s/\varepsilon_0$ smoothly. At the tricritical point the curve of $\varepsilon/\varepsilon_0$ approaches $\varepsilon_s/\varepsilon_0$ with a vertical slope, this arises because of the tricritical exponents. For higher pressure, in the first order region, the curves are discontinuous, what is shown here is the point where the ordered phase becomes unstable – there is another, lower, temperature where the paramagnetic phase is unstable and there is a mixed phase between these points [5]. The pressure dependence of the transition temperature is clear from the temperatures at which the curves terminate for different values of $\varepsilon_s$. Since $T_c(\varepsilon)$ is a strong function of $\varepsilon$ the reduction of $\varepsilon$ in the ordered phase acts to increase $T_c$ which in turn increases $M$ as is seen from equation (3).

The magnetisation is given by plots shown in figs. 4 for the second order and tricritical regimes.

These plots include the result from the linear theory where the effects of the dependence of the transition temperature on strain are approximated by a term that is linear in $(\varepsilon - \varepsilon_s)$ which has the effect of changing the Landau parameter from $B$ to $B'$ is shown as is the result labelled linear theory, for completeness we show the plot where $T_c$ has been reduced but the value of $B$ is unchanged. The magnetisation is shown for various pressures in fig. 5.

In these figures the temperature is defined in units of the transition temperature of the unstrained material. Close to a QCP, at $\varepsilon_o$, the rapid dependence of the strain on the magnetisation means that the linearised theory is only a good approximation very close to the phase transition in the second order regime.

**III Conclusions**

We have shown that the linear theory, which is defined to be valid just near the transition, gives results that deviate substantially from the self consistent calculation at low temperatures in the second order regime and in the first order regime. We have shown that the effects of the coupling to the lattice are strongest in the region of the transition but then are reduced at lower temperatures because the magnetostrictive strain takes the system away from quantum criticality. The results at low temperatures are in strong contrast to those obtained from those obtained from a perturbation theory in which the Landau parameter $B$ is reduced as this gives a much larger value of the magnetisation at low temperatures as was seen in figs. 2a and 2b. This difference would also apply to any other mechanism for obtaining a quantum critical point which resulted in an enhanced, but constant, value of $B$.

References

1. P. Gegenwart, Q. Si and F. Steglich Nature Physics 4, 186 (2008)

2. W. Yu, F. Zamborszky, J.D. Thompson, J.L. Sarrao, M.E. Torelli, Z. Fisk and S.E. Brown, *Phys Rev Lett* **92**, 086403 (2004)

3. M. Uhlarz, C. Pfleiderer and S.M. Hayden *Phys Rev Lett* **93**, 256404 (2004)

4. Y.J. Uemura et al Nature Physics 3, 29 (2007)

5. G.A. Gehring Euro Phys Lett 82, 60004 (2008)

6. C. Pfleiderer, G.J. McMullan, S.R. Julian and G.G. Lonzarich. *Phys Rev B* **55**, 8330-8338 (1997)

Figure Captions

Fig. 1. (Color on line) The change in the strain with magnetisation. The QCP occurs for the strain $\varepsilon_0$ and the strain in the absence of magnetization is $\varepsilon_s$ and the magnetization is normalised by $M_0^2 = \dfrac{AT_0}{B}$.

Fig. 2. (Color on line) The change in the transition temperature with magnetization caused by the change in the strain.

Fig. 3. (Color on line) The strain in the magnetic phase from equation (4), the self consistent theory, the tricritical point occurs for $\varepsilon_s/\varepsilon_0 = 0.9$.

Fig. 4a and 4b. (Color on line) The magnetisation as a function of temperature in (a) the second order region and (b) at the tricritical pressure; a comparison of the linear and self consistent theory.

FH-02

Fig. 5 (Color on line) The magnetization as a function of temperature for different pressures.

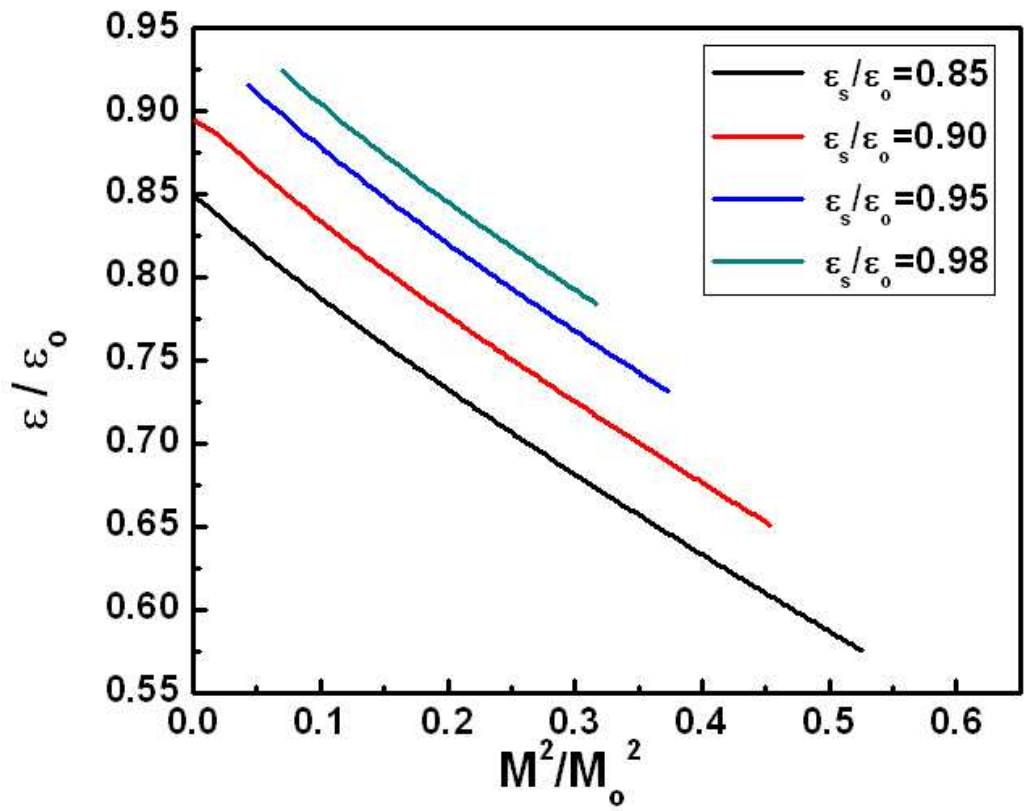

Figure 1.

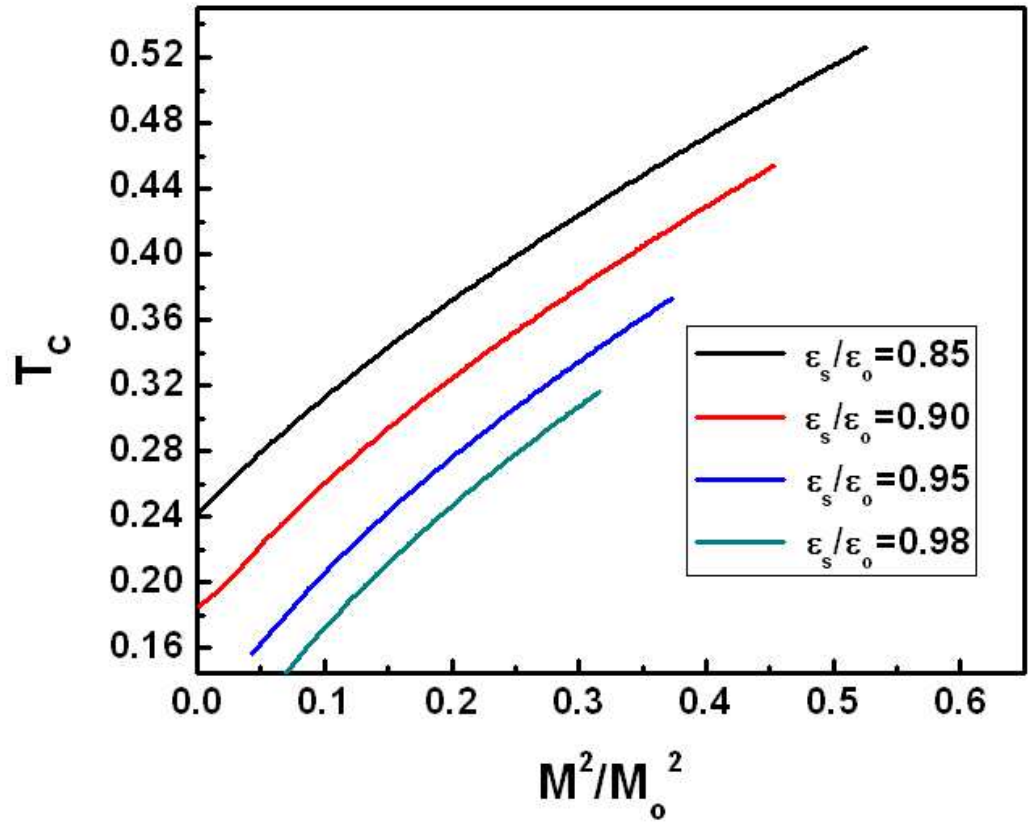

Figure 2

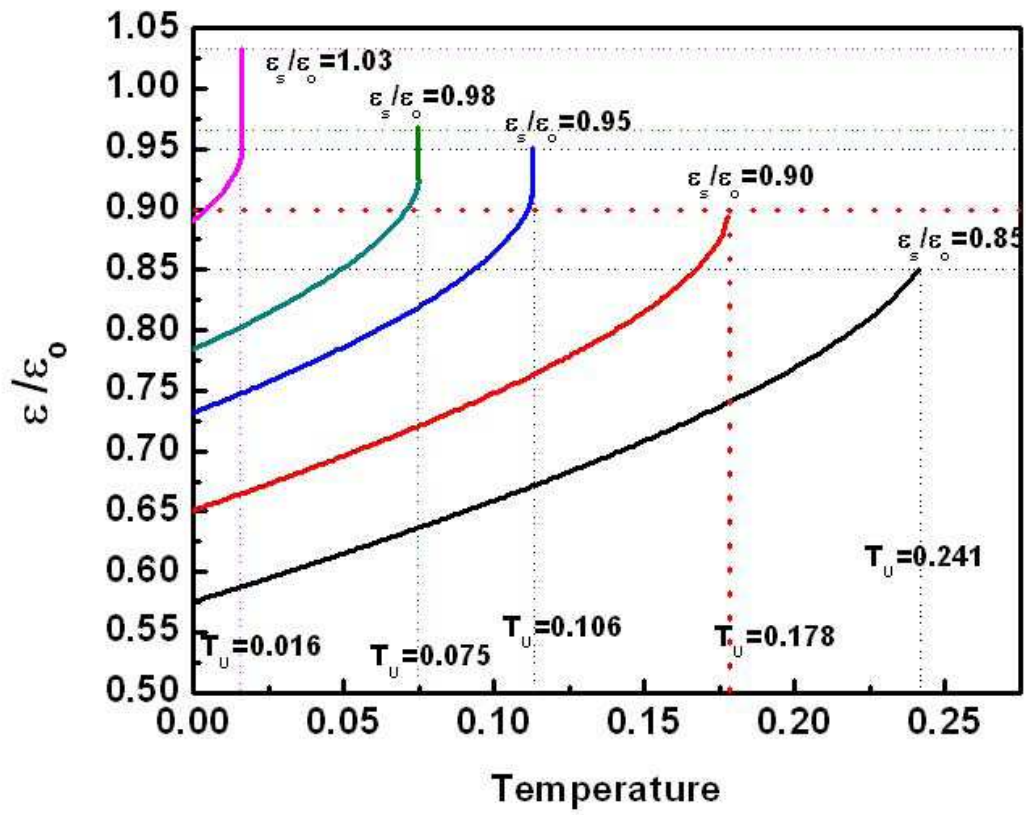

Figure 3

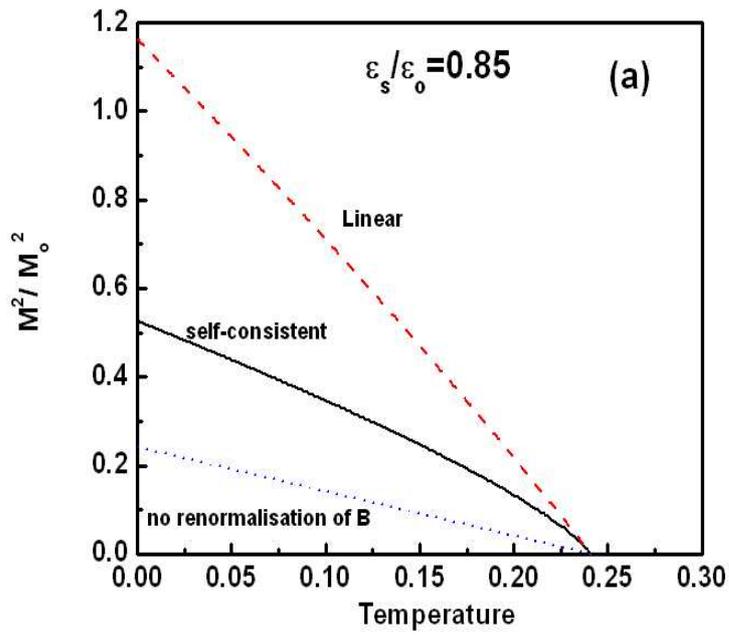

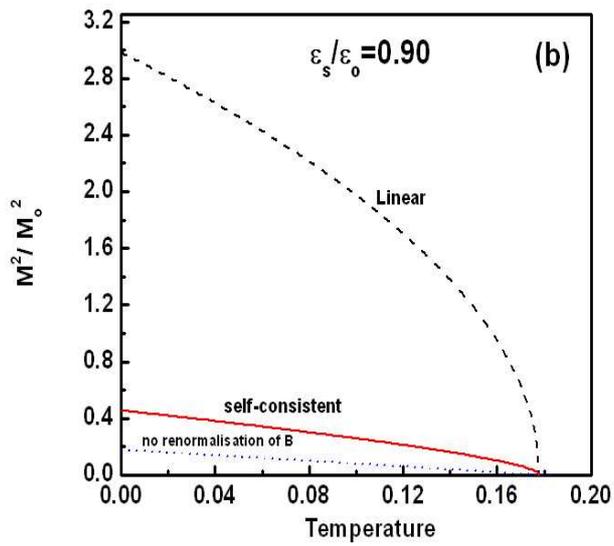

Figure 4

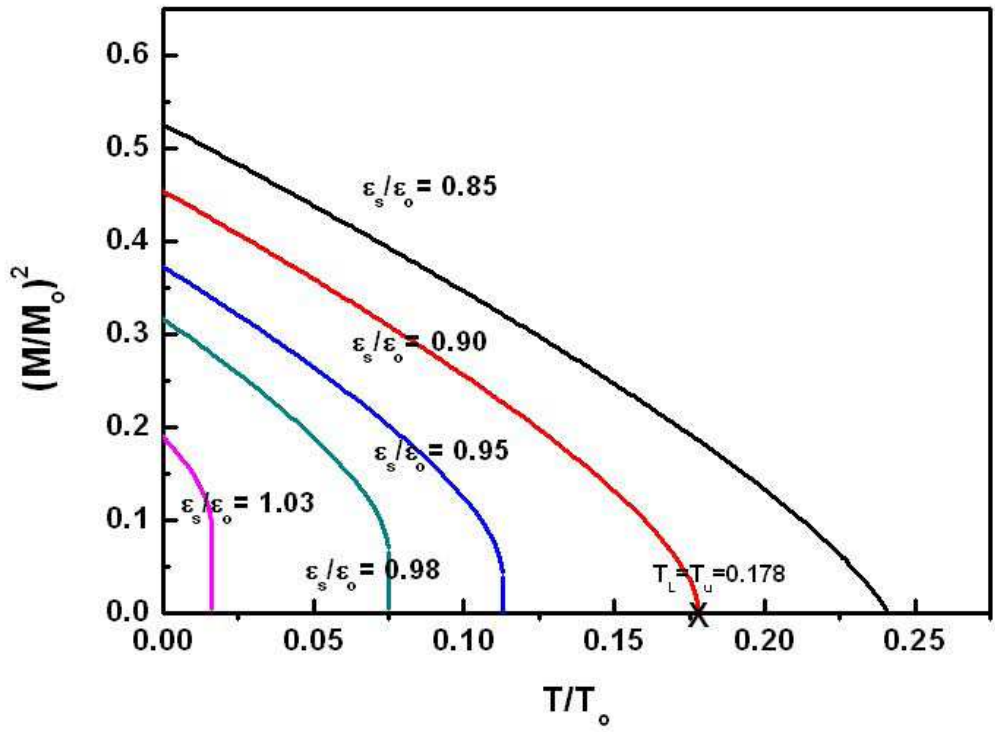

Figure 5